# Exploring the non-equilibrium fluctuation relation for quantum mechanical tunneling of electrons across a modulating barrier

Dibya J. Sivananda, Nirmal Roy, P. C. Mahato, S.S. Banerjee[*]

Department of Physics, Indian Institute of Technology Kanpur, Kanpur 208016, India.

*Email: satyajit@iitk.ac.in

## Abstract

We experimentally explore the phenomenon of electron tunneling across a modulated tunneling barrier which is created between an STM tip and an Au film deposited on a vibrating piezo surface. Measurements of the time series of the quantum mechanical tunneling current across the modulating barrier show large fluctuations. Analysis of the average work done in establishing tunneling current in finite time interval shows a distribution of both positive and negative work events. The negative work events suggest tunneling against the bias voltage direction. We show that these distributions obey the Gallavotti Cohen Non-equilibrium Fluctuation Relations (GC-NEFR) valid for systems driven through a dissipating environment. Typically, while the GC-NEFR has been shown for non - equilibrium classical systems we show its validity for the quantum mechanical tunneling process too. The GC-NEFR analysis also gives us a way to measure the dissipation present in this quantum tunneling system. We propose the modulated barrier behaves like a lossy scattering medium for the tunneling electrons resulting in a tendency to randomize the tunneling process.

## Introduction

A phenomenon which distinguishes between classical from quantum behaviour is the phenomenon of quantum mechanical tunneling of particles across a finite potential barrier. While quantum tunneling across a static barrier is a popular textbook level problem [1], the study of tunneling across a periodically modulated barrier is rich and complex. The phenomenon of tunneling across a periodically modulated barrier came into focus with experiments showing the ionization of a neutral atom due to tunneling when placed in an alternating electric field [1,2]. It was found that the tunnel ionization probability depends on both the frequency and amplitude of the drive. The tunneling across a modulated barrier is also seen in photon assisted tunneling phenomenon in Superconductor - Insulator – Superconductor junctions [3] and in quantum dots [4]. In the simple one-dimensional case, for a particle with total energy $E$ described as a plane wave, incident on a static rectangular barrier of height $V_0$ of width $w$, the tunneling probability exponentially decays with $w$ and $\sqrt{\frac{2m}{\hbar^2}(V_0 - E)}$. Within the wave picture of the tunneling particle it has been suggested that for the periodically modulated barrier, a part of the wavefunction tunnels across the barrier while a part of it remains back within the barrier. The part which is left back evolves within the modulated barrier and part of it tunnels across the periodically modulated barrier [5]. Typically for a barrier modulated at a frequency $\Omega$, the solutions for the reflected and transmitted waves not only have the usual stationary wave solution, but also have waves which are reflected and transmitted at frequency $\Omega$ [2,5,6,7]. It has also been suggested that under certain conditions, a continuous drive may lead to localization which destroys coherent tunneling across the barrier [8,9,10]. Apart from the above issues it may be noted that periodically driven quantum systems are rarely studied in isolation, as they are continuously interacting with the environment like a thermal bath. Such situations lead to dissipation effects which affect the dynamics of the evolution of the wave function in such periodically driven systems [5,11,12,13]. Such open quantum systems can no longer be studied using the conventional Schrodinger's equations [14,15,16,17] and have led to the study of interesting new phenomena like dissipative phase transitions in these open systems [18,19,20,21,22,23]. While dissipation is important in these systems, yet getting a measure of the dissipation in these systems is not clear. All these described features make a dissipative driven tunneling system interesting and rich from the point of view of exploring its nonlinear and out-of-equilibrium features, and therefore important for more experimental investigations. It also turns out that experimental realization of these systems is quite sophisticated and complex.

Fluctuation theorems offer a way to quantitively explore systems driven far from equilibrium. For a system driven through a medium where it is dissipating and exchanging energy with the surrounding (for a steady state flow), Gallavotti Cohen derived a fluctuation theorem [24,25,26,27].

$$Lt_{\tau \to \infty} \frac{P(+s_\tau)}{P(-s_\tau)} = e^{\tau s_\tau} \qquad (1)$$

Here, $s_\tau = \frac{1}{\tau}\int_t^{t+\tau} s(t')dt'$ where $s(t)$ denotes the rate at which entropy is produced in the non-equilibrium steady state. Here $P(+s_\tau)$ is the probability of entropy production, namely, of observing entropy-increasing events of magnitude $+s_\tau$, and $P(-s_\tau)$ is the probability of observing an entropy decreasing event of magnitude $-s_\tau$ where $-s_\tau$ is the entropy consumed over a duration $\tau$. Here $\tau$ is the observation time interval. Note that for relatively small observation times $\tau$, the finiteness of $P(-s_\tau)$ suggests that one observes entropy decreasing events which is a violation of the second law of thermodynamics, within the small time interval of observation. However, for large observation time interval $\tau$, as expected (see eqn.1) the probability of an entropy-increasing event $+s_\tau$ dominates over the $P(-s_\tau)$. For non - equilibrium systems, the result of exchange of energy between the system and the environment results in the entropy of the system (within a finite time window) to either increase (the usual case) or even decrease.

Different experiments have shown the validity of fluctuation relations in classical systems like dragging of a Brownian particle in an optical trap, electrical circuits, RNA stretching, Rayleigh-Bernard convection [28,29], pressure fluctuations on the surface kept in a turbulent flow [30], vertically shaken granular beads [26], Lagrangian turbulence on a free surface [31], liquid crystal electro-convection [32], vortices in superconductors [27], driven levitating nanoparticles [33], etc. While most of the studies exploring the Gallavotti Cohen Non-Equilibrium Fluctuation Relation (GC-NEFR) have been in classical systems, there have been few studies on exploring the validity of GC-NEFR in non-equilibrium quantum dissipating systems. In this work, we explore the tunneling across a modulated barrier using Scanning Tunneling Microscope (STM). Recently we have shown [34] that tunneling between the STM tip and a conducting surface (Au film) on a piezo vibrating at a frequency ω, produces a modulation in the tunneling current at frequency ω. The modulation in the tunneling current is effectively due to modulation in the tunneling barrier at frequency ω. In this paper we explore the fluctuations in the long time series of the tunneling current. We observe the presence of large excursions in the tunneling current which are both above and below the mean tunneling signal. We show that these distributions obey the well-known Gallavotti Cohen Non-Equilibrium Fluctuation Relation (GC-NEFR) valid for systems driven through a dissipating environment (steady state case). Typically, the validity of the GC-NEFR has usually been shown for non-equilibrium classical systems. The GC-NEFR analysis gives a way to measure the dissipation present in this quantum tunneling system. We propose the modulated barrier behaves like an inelastic scatterer for the tunneling electrons resulting in the observed features.

## Experimental details

We use a Quazar Technologies make room temperature STM (NanoRev. 4.0). In fig. 1(a) we show the atomic arrangement in Highly Ordered Pyrolytic Graphite (HOPG) sample imaged using this STM. We also show in fig. 1(a) the schematic representation of our STM. This set up is used to measure the time series of the tunneling currents between the STM tip and a vibrating surface. Placed below the STM tip is a conducting gold film deposited on top of a vibrating piezoelectric crystal which has a diameter ~ 1.4 cm and thickness ~ 0.33 mm. The piezo is stuck to a glass substrate (1cm × 2cm) which is then stuck to the gold-coated metallic stub of the STM with double-sided adhesive tape (fig. 1(a)). The STM circuit is completed by shorting the top conducting surface of the piezoelectric crystal with the gold-coated STM metallic stub. For our STM tip, we use an electrochemically etched Pt-Ir alloy wire and maintain a constant dc bias $V_b$ = -1.5 V between the tip and the gold film on top of the vibrating piezo surface. The STM in our experiment is operated in constant current mode. When the piezo crystal surface vibrates with frequency $f$, the tunneling gap between the STM tip and the conducting surface on top of the piezo also gets modulated with frequency $f$ (see schematic in fig. 1(a)). This results in periodic variations of the tunneling current predominantly at $f$ [34] (along with some higher harmonics). These tunneling current modulations can be measured from the voltage drop across a 1 GΩ resistor ($V_T = I_{TC} \times 10^9$ Ω) using a DSO (Digital Storage Oscilloscope). It may be mentioned here that in the constant current mode of STM operation, technically it is often easier to measure the modulations by measuring the time series of the feedback signal ($V_{FB}$). The feedback signal is a fraction of the tunneling signal ($V_T$), which is used to control the STM. We confirm this in fig. 2, which shows a linear relationship between the measured $V_{FB}$ and $V_T$ showing that the $V_{FB}$ is proportional to the tunneling signal. The linear relationship shows that the $V_{FB}$ is also a faithful representation of the tunneling signal. In our measurement, the time series of the feedback signal ($V_{FB}(t)$) from the STM is measured using a DSO (Yokagawa DL 9000 series) with a sampling rate of 5 Giga-samples per second. We would like to mention that in order to avoid any AC electrical coupling between the bottom surface of the piezo (an oscillating voltage is applied to the bottom surface to vibrate the piezo) with the piezo's top surface, the top surface of the piezo crystal (fig. 1(a)) is grounded through a 10 $\mu$F polar capacitor. The time-series signal ($V_{FB}(t)$) was captured for different frequencies $f$ of the vibrating piezo surface, ranging from 100 kHz to 1000 kHz. We would like to mention that the frequency response of our STM circuit shows significant attenuation beyond 10 MHz (see fig.5(c) of Ref.[34]). In this paper we explore a tunneling current modulation frequency range of 100 kHz to 1000 kHz, which are well within the bandwidth limitations of our STM electronics. Initially, before capturing the data, the piezo is not vibrated. In this condition, the STM is first set up to obtain an average finite dc $V_{FB}$ level which is established due to tunneling. The modulations of the feedback voltage due to the vibrating piezo below the STM tip occur around this mean dc voltage level. After this, the piezo surface is vibrated by applying an AC signal to the piezo of amplitude 30 Volts. All measurements are performed under ambient conditions.

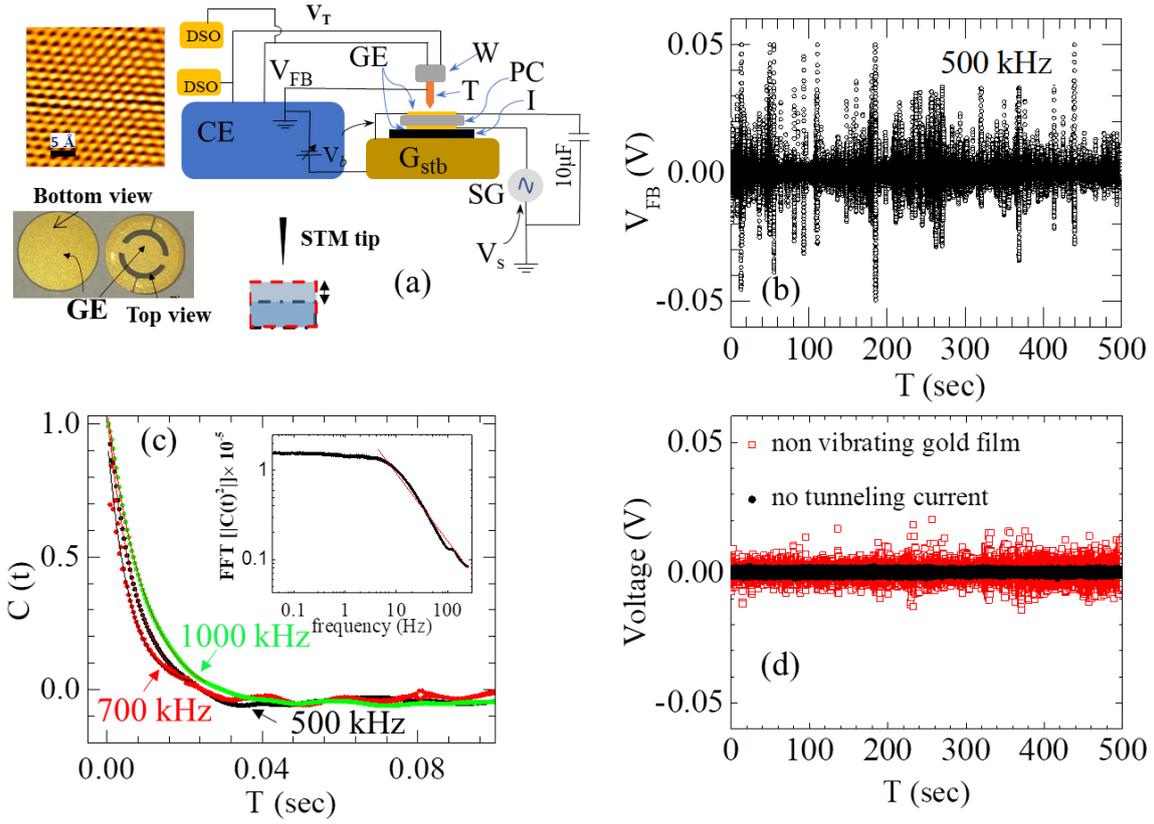

Figure 1(a): Atomic resolution image of a HOPG surface captured with our STM, schematic of the STM circuit and the front and backside of the piezoelectric crystal used. The vibration mode of the piezoelectric crystal is also shown as a schematic. The abbreviations used in the schematic are GE- Gold Electrodes; CE- STM control electronics; DSO- Digital storage oscilloscope; W- STM piezo walker; T- STM tip; PC-piezo crystal below tip; I- insulating layer; SG –function signal generator; $G_{stb}$- Gold coated stub; $V_{FB}$ – feedback signal of the STM; $V_T$ – tunneling voltage signal; $V_b$– tip to sample bias voltage; $V_s$- voltage signal from signal generator. (b) The time series of the feedback voltage ($V_{FB}$) of the STM at 500 kHz vibrating frequency of the piezoelectric crystal (c) The auto-correlation function for the feedback voltages at different frequencies of the vibrating piezo. Inset shows the $\frac{1}{f}$ noise of the 500 kHz signal (shown in fig. 1(b)) in log-log scale. (d) The electronic noise of the STM with both the tunneling current on and off. It shows that electronic noise is one order of magnitude smaller than the feedback signal.

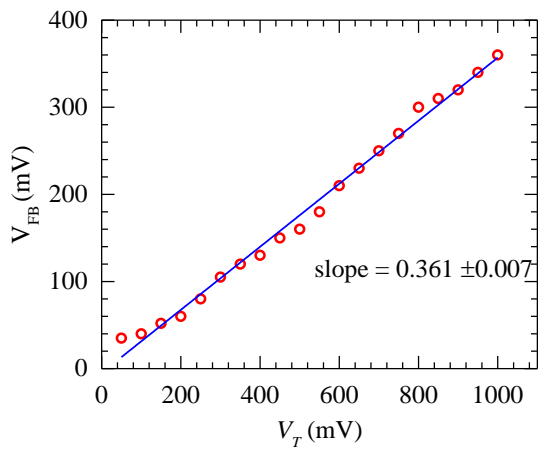

Figure 2: Shows the variation of the feedback signal with the tunneling voltage in the STM.

Figure 1(b) shows the time series of the feedback voltage for piezo excitation frequencies ($f$) of 500 kHz captured for a time duration of 500 seconds (note the signal shown is obtained after subtracting the mean dc $V_{FB}$ level discussed above). The time-series data of the modulating feedback tunneling voltage $V_{FB}(t)$ captured over tens of microseconds time interval has been explored in ref. [34]. Figures 1(b), 3(a), (d) and (g) show long time-series data of $V_{FB}(t)$ spanning few hundreds of seconds. Here we see large fluctuations which are associated with fluctuations in the tunneling current. Figure 1(d) (red data points) shows the intrinsic electronic noise in $V_{FB}(t)$ signal when the tip is positioned over a non-vibrating thin Au film. It clearly doesn't show any of the large fluctuations observed in fig.1(b). Also, in fig.1(d) we show with the black data points, the time series of the bare STM electronics noise floor voltage signals when there is no tunneling current. Here too, we do not observe large fluctuations like those of fig.1(b). We must mention that all experiments discussed here are performed under identical conditions. While the fluctuations in the $V_{FB}(t)$ from a non-vibrating substrate are more than the bare electronic noise floor, they are still an order of magnitude less than that with a vibrating quartz surface below the STM tip (fig. 1(b)).

From the time series of $V_{FB}(t)$ data (e.g. fig 1(b)) captured for different piezo excitation frequencies (500 kHz, 700 kHz and 1000 kHz) between the STM tip and the vibrating piezo surface we calculate the autocorrelation function using the expression $C(t) = \frac{<V_{FB}(t+t') V_{FB}(t)>}{V_0^2}$ where $V_0$ is the mean of $V_{FB}(t)$ and $<..> = \frac{1}{T}\int_0^T ..dt'$ (see fig. 1(c)). The $C(t)$ shows the $V_{FB}(t)$ signals are uncorrelated beyond tens of millisecond. Inset of fig. 1(c) shows the behaviour of the Fourier transform of $|C(t)|^2$ of 500 kHz data in the main panel i.e. fig. 1(c) shows the power spectrum $P(f)$ as a function of frequency on a log-log scale. Inset of fig. 1(c) shows that the power spectrum of the noise has a non-shot noise type behavior, viz. $P(f) \propto \frac{1}{f^\alpha}$ where $\alpha = 0.7 \pm 0.1$ (recall that Flicker noise is characterized by $P(f) \propto \frac{1}{f}$). The absence of shot noise features suggest that the fluctuations are not related to the discreetness of the tunneling of electronic charges.

In fig. 3(b) we analyze the time series in fig. 3(a) in terms of $P(W_\tau)$. Here the $P(W_\tau)$ is the probability of observing an event of magnitude $W_\tau$ within an observation time interval of $\tau$. From the $V_{FB}(t)$ signal (after subtracting the dc offset voltage), like the ones shown in fig. 3(a), the feedback voltage time series $V_{FB}(t)$ is broken up into a series of time bins each of width $\tau$ where we calculate $W_\tau$ as

$$W_\tau = \frac{s_\tau}{<s(t)>} = \frac{\frac{1}{\tau}\int_t^{t+\tau} I_T(t')V_b\, dt'}{V_b\langle I_T\rangle} \qquad (2)$$

Here $I_T$ is the tunneling current, and $V_b$ is the constant dc STM bias voltage applied between the STM tip and base. Note $I_T \propto V_T$ where $V_T$ is the tunneling voltage and as $V_{FB} \propto V_T$ (fig. 2). Therefore,

$$W_\tau = \frac{s_\tau}{<s(t)>} = \frac{\frac{1}{\tau}\int_t^{t+\tau} V_{FB}(t')V_b\, dt'}{V_b\langle V_{FB}\rangle} \qquad (3)$$

Where $<V_{FB}>$ is the average value of the $V_{FB}(t)$ signal within an observation time interval $\tau$. $W_\tau$ is the average work done in the binning time width $\tau$, for electrons tunneling from the STM tip onto the vibrating surface of the piezo.

Often for experimental studies, the transformed form of the Gallavotti Cohen Non-equilibrium fluctuation relation (GC-NEFR) is used, where the GC-NEFR of eqn.1 is restated in terms of $W_\tau$ [25,26]

$$R = \frac{1}{\tau}\ln\left(\frac{P(+W_\tau)}{P(-W_\tau)}\right) = s_\tau = W_\tau <s(t)> \qquad (4)$$

Here, $s(t) = IV/k_B T_{eff}$ [25,26,35], where $IV(t)$ is the instantaneous power flux into the system and $T_{eff}$ is the effective temperature of the system. Here $T_{eff}$ is an effective temperature scale for non – equilibrium systems, and it has no connections with the equilibrium temperature. Experimental determination from different driven situations have shown $T_{eff}$ values are typically high and they vary between $10^6$ to $10^{16}$ K depending on the non-equilibrium situation being studied [35,25,26,27,35]. Here the time scale $\tau$ is the time window bins over which the time series data is integrated. Note that any experiment searching for validity of the GC-NEFR as per equation 4, should observe a linear relationship between $R$ vs $W_\tau$ with a unique slope whose value should not depend on the choice of $\tau$.

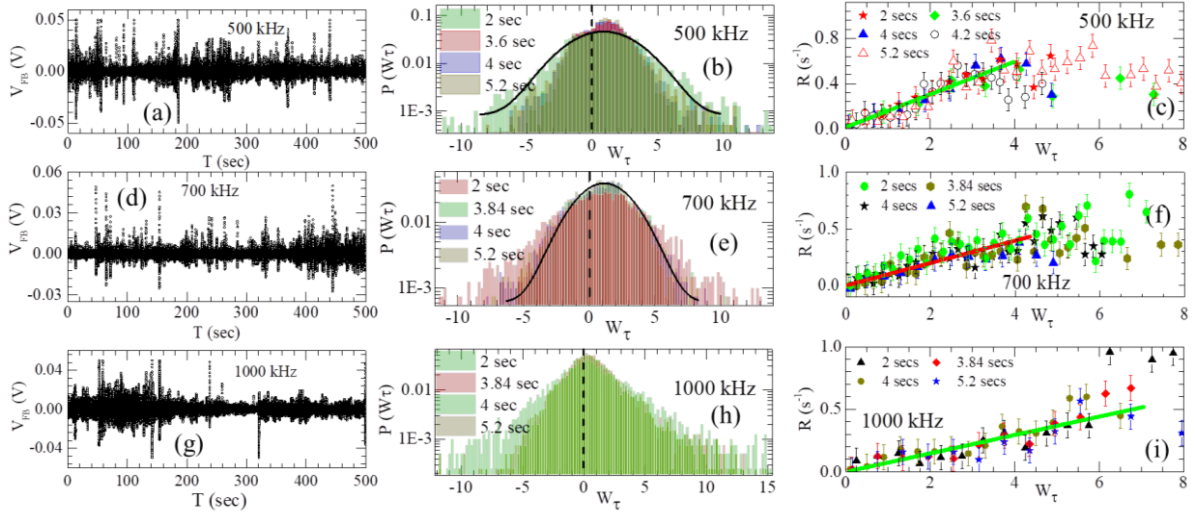

Figure 3:(a) Time series of the feedback voltages for a frequency of 500 kHz of the vibrating piezo. (b) The probability distribution functions (PDF) of the time series in (a). (c) The $R$ v/s $W_\tau$ for 500 kHz frequency. (d) Time series of the feedback voltages for a frequency of 700 kHz of the vibrating piezo. (e) The probability density functions of the time series in (d). (f) The $R$ v/s $W_\tau$ for 700 kHz frequency. (g) Time series of the feedback voltages for a frequency of 1000 kHz of the vibrating piezo. (h) The probability density functions of the time series in (g). (i) The $R$ v/s $W_\tau$ for 1000 kHz frequency.

We determine $P(W_\tau)$ from the time series signal in fig. 3(a) and plot it in fig. 3(b) for different choice of $\tau$. From fig. 3(b) we see that the $P(W_\tau)$ versus $W_\tau$ curve, also called the Probability Distribution Function (PDF), has a non-Gaussian distribution around a positive mean peak of $<W_\tau> \sim 1.1$. We see from the $P(W_\tau)$ distribution that it not only has positive work events but also has a significant probability of negative work, $-W_\tau$ events. The $-W_\tau$ events arise from the excursions of $V_{FB}(t)$ signal which fall below the mean $V_{FB}$ signal as noted in fig.1(b) earlier. The positive $W_\tau$ events represent entropy increasing events viz. work done within observation time interval $\tau$ when tunneling is established along the bias voltage. The negative, $-W_\tau$ event are unusual events, representing the average work done against the bias drive within time interval $\tau$. For these negative work events it appears as if, within a finite observation interval $\tau$, one catches glimpses of tunneling electrons which seem to be swimming up against the bias voltage. All of the above features which are seen in fig. 3(b) for 500 kHz excitation of the vibrating piezo surface are reproduced at other excitation frequencies of 700 kHz and 1000 kHz as well (see figs. 3(e) and 3(h)).

Figure 3(c) shows a plot of $R$ vs $W_\tau$ for different choices of $\tau$ for the 500 kHz data. We see a linear relation between $R$ and $W_\tau$ for $0 < W_\tau \leq 4$. For different choice of $\tau$, all the $(W_\tau, R)$ data scale onto a single curve (for comparison the unscaled feature of the data is shown in supplementary [36]), which is consistent with eqn.(4). The best fit curve through the data points within the error bars, is a straight line with a unique slope (shown as the green solid line in fig.3(c)). This feature is also consistent with eqn.(4). For higher $W_\tau$, the scaling feature and the linear relation between $R$ and $W_\tau$ completely break down and a significant dependence on $\tau$ is seen. Figures 3(f) and (i) shows similar behavior for the 700 kHz and 1000 kHz data. Infact for 1000 kHz (fig.3(i)) the linear relationship between R and $W_\tau$ extends upto $W_\tau \sim 6$. Thus, the observation of scaling behavior of $(W_\tau, R)$ and the best fit curve is a straight line through the scaled data with a unique slope, shows that eqn.(4) is validated and GC-NEFR is valid within within $0 < W_\tau \leq 4$. Note that other experiments on classical driven systems have validated the GC-NEFR (eqn.4) within a limited range of $W_\tau$ variation, by using similar scaling and linear fitting of data [25,26,27,33,35].

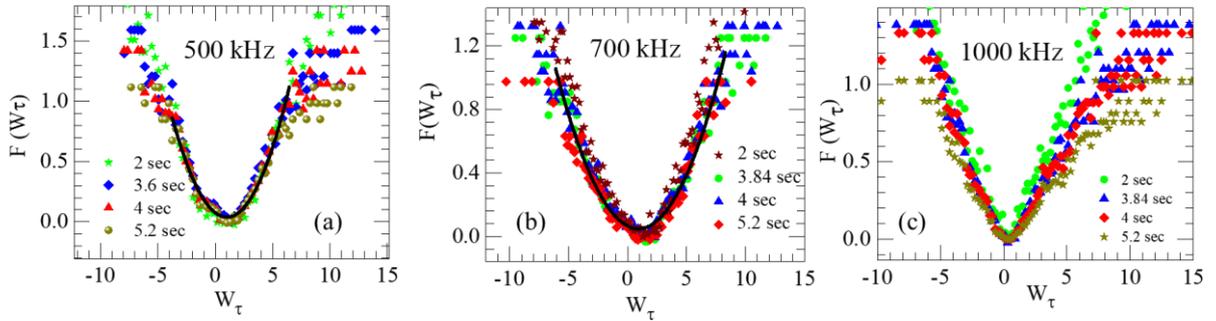

Figure 4: $F(W_\tau)$ curves for different $\tau$ for (a)500 kHz (b) 700 kHz and (c) 1000 kHz of the vibrating piezo. Solid (black) line represents the data fitted with a form $a_0(W_\tau - \langle W_\tau \rangle)^2$, where the constant $a_0$ is 0.033 (500 KHz), 0.020 (700 kHz) and $\langle W_\tau \rangle$ = 1.1 for 500 kHz and 700 kHz. At 1000 kHz the $F(W_\tau)$ cannot be fitted to a quadratic form.

For a large observation time interval, for driven systems, the Large deviation function (LDF) is a measure of the probability of observing events with a large value of an observable (viz., events with values larger than the mean of the observable) [37,38]. The non-equilibrium fluctuation relation is related to the symmetry property of the LDF [37,38]. To analyze the shapes of the $P(W_\tau)$ curves in figs. 3(b), 3(e) and 3(h) we plot the LDF function, $F(W_\tau)$ [25,27,37,38].

$$F(W_\tau) \propto - ln[P(W_\tau)] \qquad (5)$$

If $P(W_\tau)$ has a Gaussian form then $F(W_\tau) \propto (W_\tau - \langle W_\tau \rangle)^2$ viz. $F(W_\tau)$ has a quadratic dependence on $W_\tau$. Using the $P(W_\tau)$ data of fig. 3(b) and eqn. 5 we determine $F(W_\tau)$ by taking the natural logarithm of $P(W_\tau)$, (see fig. 4(a)) for excitation frequencies of 500 kHz of the vibrating piezo (note our $\tau$ values are larger than the time interval over which C(t) →0). The solid line in figure 4(a) shows $F(W_\tau)$ behaves as $(W_\tau - \langle W_\tau \rangle)^2$ for $|W_\tau| \leq 5$. Note that for large $W_\tau$ (> ±5) the $F(W_\tau)$ is non-quadratic in nature. Similar feature is seen for 700 kHz data in fig 4(b). At 1000 kHz we see the $F(W_\tau)$ is completely non-quadratic in shape for all $W_\tau$, however it still obeys GC-NEFR quite well (eqn. 4). A deviation of $F(W_\tau)$ from quadratic nature represents that the fluctuations in $W_\tau$ are not random events. Thus, it seems that the fluctuations in the tunneling current through the modulated barrier may be correlated.

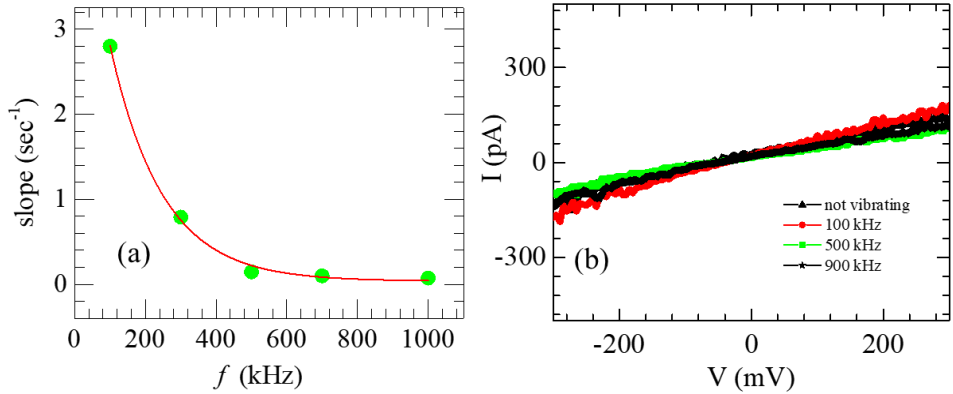

Figure 5(a): Shows the variation of slope ($\propto \frac{1}{T_{eff}}$) with the frequency of the vibrating piezo. The solid line is a guide to the eye. (b) *I-V* while tunneling onto the gold conducting surface on the piezo for different frequencies of vibration.

It may be also noted that 1000 kHz is near resonance of the piezo. The deviation from Gaussian could be related to additional dissipation induced in the quantum tunneling occurring in the system near resonance. In the GC-NEFR, in the curve between $R$ v/s $W_\tau$ in eqn. (4), the slope is $\propto \frac{1}{T_{eff}}$. We use the slope of $R$ v/s $W_\tau$ curve as a measure of inverse of the dissipation ($\delta^{-1}$) in this driven quantum tunneling system. In fig. 5(a) we plot the slope ($\propto \delta^{-1}$) v/s the frequency of the piezo crystal's vibration. It is clear that beyond 400 kHz of modulation frequency of the barrier, the dissipation in the tunneling quantum system increases. We show in fig. 5(b) that the local *I-V* of the Au film (on top of the piezo) measured with the STM for different vibrating frequencies are identical, suggesting that the states of the Au film into which the electrons are tunneling into, are identical at different frequencies. Hence, any change in the density of electronic states of the film is not responsible for the observed features.

Although at present we don't understand the mechanism of dissipation, we envisage the following scenario. It is known that a statistical description of the behavior of the tunneling current for a modulating barrier can be obtained by considering a part of the wave tunneling through and a part remaining trapped within the barrier [5]. However, this is not sufficient to explain the dissipation and negative entropy events observed in our experiments. We believe the modulated barrier should possess internal quantized energy states associated with the part of the electron wave trapped within the barrier from prior tunneling events. With the modulating barrier (due to the changing STM tip to sample distance), the barrier energy level spacings are also periodically modulated. We believe that an electron impinging on the modulated barrier, should scatter inelastically. Due to the modulating energy levels, it is possible that the impinging tunneling electron may cause resonant excitations within levels of the barrier and lead to losses in energy from the tunneling electron which go into causing excitations within the barrier. Some electrons may also be reflected back along with energy loss, and some may tunnel across with energy loss. The reflected electrons may account for the negative fluctuation events observed. Hence for quantum tunneling across a modulated tunneling barrier between the STM tip and piezo surface, the barrier at different frequencies behaves like a lossy medium, which appears to become impervious to some of the tunneling electrons while some tunnel through. The loss of this medium is a function of frequency.

In conclusion we show the validity of the Gallavotti Cohen Non-equilibrium fluctuation relation for a dissipating quantum tunneling system. We believe the analysis provides a useful way to quantify the dissipation in this quantum system. The validity of the GC-NEFR shows that the symmetries which govern the classical GC non-equilibrium fluctuation relations are valid in the quantum regime too. More future theoretical and experimental investigations are needed to understand the complexity of electron tunneling across a modulated barrier.

*Acknowledgement*

S. S. B. acknowledges discussions with Prof. Ajay Sood of Dept of Physics, IISc Bangalore. S.S.B would like to acknowledge funding support from IITK (IN) and DST-TSDP (IN) DST-SERB Imprint II (IN) Government of India.

# Supplementary information

-------------------------

# Exploring the non-equilibrium fluctuation relation for quantum mechanical tunneling of electrons across a modulating barrier

Dibya J. Sivananda, Nirmal Roy, P. C. Mahato, S.S. Banerjee[*]

Department of Physics, Indian Institute of Technology Kanpur, Kanpur 208016, India.

*Email: satyajit@iitk.ac.in

-------------------------

**Non scaling feature of the ratio of probability of power consumption determined for different choice of $\tau$ , and the appropriate quantity which scaling the data**

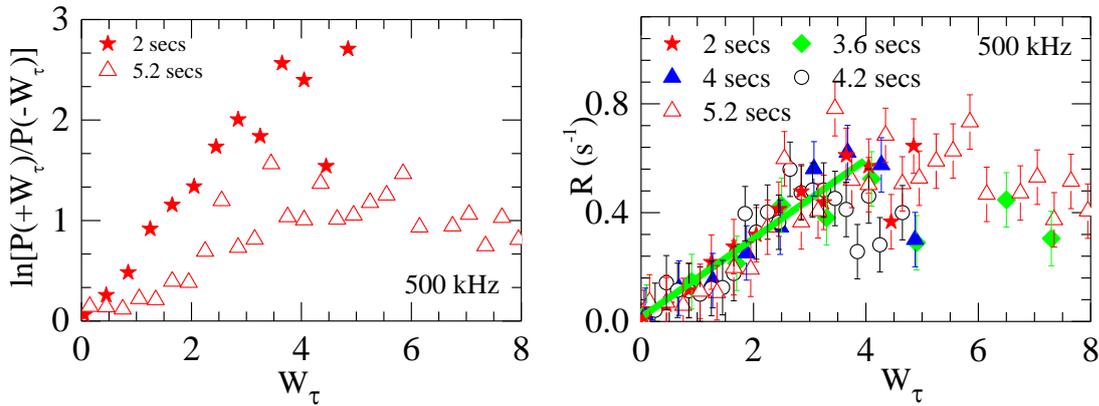

Recall equation (4) in the MS

$$R = \frac{1}{\tau} ln \left( \frac{P(+W_\tau)}{P(-W_\tau)} \right) = s_\tau = W_\tau <s(t)> \qquad (4)$$

In the figure on the left for 500 kHz time series data, we plot $ln \left( \frac{P(+W_\tau)}{P(-W_\tau)} \right)$ vs $W_\tau$ for two very different values of choice of $\tau$. The figure clearly shows that the data for 2 seconds and 5.2 seconds do not overlap. The plot on the right shows that the analysis becomes independent of the choice of $\tau$ and all the data scale onto a single curve by plotting $R$ vs $W_\tau$ (which follows from eqn.4). We see that within $0 < W_\tau < 4$ , the best fit curve through the scaled data in the plot on right is a straight line with a unique slope.